# Stationary solutions of second-order equations for point fermions in the Schwarzschild gravitational field

V.P.Neznamov[1,2*], I.I.Safronov[1]

[1]FSUE "RFNC-VNIIEF", Russia, Sarov, Mira pr., 37, 607188
[2]National Research Nuclear University MEPhI, Moscow, Russia

Abstract

When using a second-order Schrödinger-type equation with the effective potential of the Schwarzschild field, existence of a stationary state of half-spin particles with energy $E=0$ is proved. For each of the values of quantum numbers $j, l$, the physically meaningful energy $E=0$ (the binding energy is $E_b = mc^2$) is implemented at the value of the gravitational coupling constant $\alpha \geq \alpha_{\min}$. The particles with $E=0$ are, with the overwhelming probability, at some distance from the event horizon within the range from zero to several fractions of Compton wavelength of a fermion depending on value of the gravitational coupling constants and values $j, l$.

In this paper, similar solutions of the second-order equation are announced for bound states of fermions in the Reissner-Nordström, Kerr, Kerr-Newman fields.

Atomic-type systems: collapsars with fermions in bound states are proposed as particles of dark matter.

[*] E-mail: vpneznamov@vniief.ru; vpneznamov@mail.ru

## 1. Introduction

In quantum mechanics, in order to describe the motion of half-spin particles, Dirac equations with bispinor wave functions are commonly used. Actually simultaneously, Dirac also proposed a second-order equation for electrons and positrons in the external electromagnetic field [1]. Using the ratio between the upper and lower spinors of the Dirac bispinor, the second-order equation can be written as two separate equations with spinor wave functions. In this case, to provide the self-conjugacy, it is necessary to perform appropriate nonunitary similarity transformations in each of the two second order equations with spinor wave functions (see, e.g., [2]). Such transformations preserve energies of quantum-mechanical states but do not preserve probabilities of states. As a result, use of self-conjugate second-order equations with spinor wave functions in quantum mechanics of half-spin particles in external electromagnetic and gravitational fields can lead to new physical consequences.

In this paper, we discuss stationary solutions of Dirac equations and the second-order equation in the Schwarzschild gravitational field [3].

Earlier, in multiple efforts, existence of nonstationary solutions of Dirac equations, corresponding to bound states of Dirac particles with complex energies exponentially decaying with time, was shown in the Schwarzschild space-time. Existence of resonance states in the Schwarzschild field for massive scalar particles was discussed in [4] - [7] by using Klein-Gordon equation. The similar problem for Dirac particles was examined in [8] - [13].

In this paper, we obtained the stationary solution of Dirac equation with the fermion energy $E = 0$ (the binding energy is $E_b = mc^2$, where $m$ is the fermion mass). However, this solution is not physical since the wave function of state with $E = 0$ is not square-integrable due to the logarithmic divergence of the normalization integral near the event horizon. In this sense, we validate the absence of physically meaningful stationary solutions of the Dirac equation in classical Schwarzschild, Reissner-Nordström [18], Kerr [19], Kerr-Newman [20] fields, proved earlier in [14] - [17].

For stationary solutions of the second-order equation, the situation qualitatively changes. The solution of $E = 0$ corresponds to square-integrable wave eigenfunctions vanishing on the event horizon. Wave functions depend on the gravitational coupling constant $\alpha$ and on quantum numbers of the angular and orbital moments of half-spin particles.

The solutions of the second-order equation, corresponding to energies of bound states of half-spin particles for Reissner-Nordström, Kerr, Kerr-Newman fields, are also presented in the paper as an announcement. These results are to be validated in the following papers.

The paper is organized as follows. In section 2, a self-conjugate Dirac Hamiltonian is presented in the Schwarzschild field with the flat (without a Parker weight factor [21]) scalar product of wave functions. The Hamiltonian was earlier obtained in [23] by methods of pseudo-Hermitian quantum mechanics used in [22] - [24] to obtain self-conjugate Dirac Hamiltonians in arbitrary gravitational fields, those depending on time. In section 2, variables are separated, equations and asymptotics for radial functions are presented.

Section 3 presents two self-conjugate second-order equations with effective potentials. Each of these equations is referred to only one of the two transformed radial wave functions. In the section, singularities and asymptotics of effective potentials are examined, asymptotics of wave functions of the second-order equation is determined, existence of the regular stationary solution with $E=0$ is proved with appropriate square-integrable wave functions.

In section 4, the Prüfer transformation [25] - [28] is implemented for the numerical solution of the second-order equation. Boundary conditions at $r \to \infty$, $r \to r_0$ ($r_0$ is the gravitational radius) are specified and the numerical method of solving the equation with the phase function $\Phi(r)$ is briefly commented.

In section 5, the results of numerical calculations are presented. For the values of $\alpha > 0.25$, the only stationary bound state of fermions is shown to exist with $E=0$. The particles with $E=0$ are, with the overwhelming probability, at the distance from the event horizon within the range from zero to several fractions of Compton wave length of a fermion depending on the value of the gravitational coupling constant $\alpha$ and the values of quantum numbers $j, l$.

In section 6, energies of bound states of half-spin particles in Reissner-Nordström, Kerr, Kerr-Newman fields are presented.

In section 7 The Schwarzschild collapsars with stationary bound Dirac particles explored as candidates for the role of the dark matter particles.

In section 8, the possibility of transition from the quantum mechanics to the classical description is discussed.

In the Conclusions, basic results of the effort are discussed.

In Appendix A and B, the procedure for obtaining and the explicit form of effective potentials of the second-order equation in the Schwarzschild field is presented.

## 2. Dirac equation in the Schwarzschild field.

The Schwarzschild solution is characterized by a point source of the gravitational field with mass $M$ and the gravitational radius (event horizon)

$$r_0 = 2GM/c^2. \tag{1}$$

Here, $G$ is the gravitational constant, $c$ is the velocity of light. For a test particle with mass $m$, the dimensionless gravitational coupling constant is

$$\alpha = GMm/\hbar c = Mm/M_P^2 = r_0/2l_c. \tag{2}$$

In (2), $M_P = \sqrt{\hbar c/G} = 2.2 \cdot 10^{-5}$ g is the Planck mass, $l_c = \hbar/mc$ is the Compton wave length of a particle.

As opposite to the coupling constants in the Standard model, the coupling constant $\alpha$ can achieve very high values. For an electron, the source of gravitation with mass $M = 0.5 \cdot 10^{15}$ kg corresponds to the value of $\alpha \simeq 1$. Then, the gravitational interaction of an electron with the source with the solar mass $M = M_\odot \simeq 2 \cdot 10^{30}$ kg is determined by the value $\alpha \simeq 4 \cdot 10^{15}$.

Below, as a rule, we use the system of units $\hbar = c = 1$, the signature of the Minkowski space-time is selected to be

$$g_{\alpha\beta} = \text{diag}[1, -1, -1, -1]. \tag{3}$$

The underlined indeces are local ones. The notations $\gamma^\alpha, \gamma^{\underline{\alpha}}$ correspond to the Dirac world and local matrices. We use matrices in the Dirac-Pauli representation as local matrices.

The Schwarzschild metric is

$$ds^2 = f_S dt^2 - \frac{dr^2}{f_S} - r^2 \left(d\theta^2 + \sin^2\theta d\varphi^2\right), \tag{4}$$

where $f_S = 1 - \dfrac{r_0}{r}$.

In [23], the stationary self-conjugate Dirac Hamiltonian with the flat (without the Parker weight factor [21]) scalar product of wave functions was obtained for the Schwarzschild metric:

$$H_\eta = H_\eta^+ = \sqrt{f_S}\, m\gamma^{\underline{0}} - i\sqrt{f_S}\,\gamma^{\underline{0}} \left\{ \gamma^{\underline{1}} \sqrt{f_S} \left( \frac{\partial}{\partial r} + \frac{1}{r} \right) + \right.$$

$$\left. + \gamma^{\underline{2}} \frac{1}{r}\left( \frac{\partial}{\partial \theta} + \frac{1}{2}\text{ctg}\theta \right) + \gamma^{\underline{3}} \frac{1}{r\sin\theta}\frac{\partial}{\partial \varphi} \right\} - \frac{i}{2}\gamma^{\underline{0}}\gamma^{\underline{1}} \frac{\partial f_S}{\partial r}. \tag{5}$$

In (5), the sign "+" means the Hermitian conjugation. The Dirac equation with Hamiltonian (5) has the form

$$i\frac{\partial \Psi_\eta}{\partial t} = H_\eta \Psi_\eta. \tag{6}$$

## 2.1 Separation of variables

Dirac equation (6) allows separation of angular and radial variables. For stationary states, it is convenient to define the bispinor $\Psi_\eta(\mathbf{r},t)$ as [13]

$$\Psi_\eta(r,\theta,\varphi,t) = \begin{pmatrix} F(r)\xi(\theta) \\ -iG(r)\sigma^3\xi(\theta) \end{pmatrix} e^{-iEt} e^{im_\varphi \varphi} \qquad (7)$$

and to use the Brill-Wheeler equation [29]

$$\left[-\sigma^2\left(\frac{\partial}{\partial\theta}+\frac{1}{2}\operatorname{ctg}\theta\right)+i\sigma^1 m_\varphi \frac{1}{\sin\theta}\right]\xi(\theta) = i\kappa\xi(\theta). \qquad (8)$$

In order to use equation (8), it is necessary to perform the equivalent replacement of $\gamma$-matrices in Hamiltonian (5):

$$\gamma^1 \to \gamma^3, \ \gamma^3 \to \gamma^2, \ \gamma^2 \to \gamma^1. \qquad (9)$$

In equations (7), (8), the spinor $\xi(\theta)$ represents spherical harmonics for a half spin; $\sigma^i$ are the two-dimensional Pauli matrices; $m_\varphi$ is the azimuthal component of the angular momentum $j$; $\kappa$ is the quantum number of the Dirac equation:

$$\kappa = \mp 1, \mp 2, \ldots = \begin{cases} -(l+1), & j = l+1/2 \\ l, & j = l-1/2 \end{cases}. \qquad (10)$$

In (10), $j, l$ are quantum numbers of the angular and orbital momenta of the Dirac particle.

The spinor $\xi(\theta)$ can be represented as [13]

$$\xi(\theta) = \begin{pmatrix} -\tfrac{1}{2} Y_{jm_\varphi}(\theta) \\ \tfrac{1}{2} Y_{jm_\varphi}(\theta) \end{pmatrix} = (-1)^{m_\varphi+\tfrac{1}{2}} \sqrt{\frac{1}{4\pi}\frac{(j-m_\varphi)!}{(j+m_\varphi)!}} \begin{pmatrix} \cos\tfrac{\theta}{2} & \sin\tfrac{\theta}{2} \\ -\sin\tfrac{\theta}{2} & \cos\tfrac{\theta}{2} \end{pmatrix} \times$$
$$\times \begin{pmatrix} \left(\kappa - m_\varphi + \tfrac{1}{2}\right) P_l^{m_\varphi - \tfrac{1}{2}}(\theta) \\ P_l^{m_\varphi + \tfrac{1}{2}}(\theta) \end{pmatrix}, \qquad (11)$$

where the expression after the square root in parentheses is a two-dimensional matrix; $P_l^{m_\varphi \pm 1/2}(\theta)$ are the associated Legendre functions.

As a result of separation of variables, we obtain equations for the radial functions $F(\rho), G(\rho)$. Then, we write these equations in dimensionless variables of $\varepsilon = E/m$, $\rho = r/l_c$, $r_0/l_c = 2\alpha$, $l_c = m^{-1}$. Below, the motion of a fermion is considered relative to the resting source of the Schwarzschild field, i.e., $m \ll M$.

## 2.2 Equations and asymptotics for radial wave functions

The system of equations for the radial functions $F(\rho), G(\rho)$ has the form

$$\left(1-\frac{2\alpha}{\rho}\right)\frac{dF}{d\rho}+\left(\frac{1+\kappa\sqrt{1-\frac{2\alpha}{\rho}}}{\rho}-\frac{\alpha}{\rho^2}\right)F-\left(\varepsilon+\sqrt{1-\frac{2\alpha}{\rho}}\right)G=0,$$

$$\left(1-\frac{2\alpha}{\rho}\right)\frac{dG}{d\rho}+\left(\frac{1-\kappa\sqrt{1-\frac{2\alpha}{\rho}}}{\rho}-\frac{\alpha}{\rho^2}\right)G+\left(\varepsilon-\sqrt{1-\frac{2\alpha}{\rho}}\right)F=0.$$

(12)

The asymptotics of solutions (12):

At $\rho \to \infty$, the leading terms of the asymptotics are (see, e.g., [13])

$$F = C_1\varphi_1(\rho)e^{-\rho\sqrt{1-\varepsilon^2}}+C_2\varphi_2(\rho)e^{\rho\sqrt{1-\varepsilon^2}},$$

$$G = \sqrt{\frac{1-\varepsilon}{1+\varepsilon}}\left(-C_1\varphi_1(\rho)e^{-\rho\sqrt{1-\varepsilon^2}}+C_2\varphi_2(\rho)e^{\rho\sqrt{1-\varepsilon^2}}\right),$$

(13)

where $\varphi_1(\rho), \varphi_2(\rho)$ are power functions of $\rho$.

To ensure the finite motion of Dirac particles, it is necessary to use only exponentially decreasing solutions (13), i.e., in this case $C_2 = 0$.

At $\rho \to 2\alpha$ $(r \to r_0)$, let us represent the functions $F(\rho), G(\rho)$ as

$$F\big|_{\rho\to 2\alpha} = v^s\sum_{k=0}^{\infty}f_k v^k,$$

$$G\big|_{\rho\to 2\alpha} = v^s\sum_{k=0}^{\infty}g_k v^k,$$

(14)

where $v = |\rho - 2\alpha|$.

The indicial equation for system (12) leads to the solution

$$s = -\frac{1}{2} \pm i2\alpha\varepsilon.$$

(15)

It is seen from (15) that the oscillating parts of solutions (14) vanish at zero energy of a fermion $\varepsilon = 0$.

Formally, the solution $\varepsilon = 0$ is the unique regular solution of system (12), however, this solution does not correspond to the real physical situation due to the logarithmic divergence of the normalization integral

$$N_D = \int \left(F^*(\rho)F(\rho)+G^*(\rho)G(\rho)\right)\rho^2 d\rho$$

(16)

in the vicinity of the event horizon.

As a result, we, as well as the authors of [14] - [17], arrive at the conclusion of absence of solutions to the Dirac equation corresponding to stationary bound states of fermions in the Schwarzschild field. Below, we will examine this problem as applied to the solutions of the second-order equation with spinor wave functions.

## 3. Second-order equation for fermions in the Schwarzschild field.

To obtain a second-order equation, three stages are to be implemented:
1. obtaining a self-conjugate Hamiltonian or a self-conjugate Dirac equation;
2. transition from bispinor to spinor wave functions in the second-order equation;
3. nonunitary similarity transformation to ensure self-conjugacy of the second-order equation with spinor wave functions.

Similarity transformation ensures conservation of fermion energy at transition from the Dirac equation to the second-order equation.

The procedure of obtaining second-order equations for fermions in the Schwarzschild field is presented in Appendix A. Equations for the radial functions $\psi_F(\rho), \psi_G(\rho)$ have the form of the Schrödinger equation with the effective potentials $U_{eff}^F(\rho), U_{eff}^G(\rho)$, nonlinearly depending on the energy $\varepsilon$:

$$\frac{d^2\psi_F(\rho)}{d\rho^2} + 2\left(E_{Schr} - U_{eff}^F(\rho)\right)\psi_F(\rho) = 0, \quad (17)$$

$$\frac{d^2\psi_G(\rho)}{d\rho^2} + 2\left(E_{Schr} - U_{eff}^G(\rho)\right)\psi_G(\rho) = 0. \quad (18)$$

In (17), (18),

$$E_{Schr} = \frac{1}{2}\left(\varepsilon^2 - 1\right), \quad (19)$$

$$\psi_F(\rho) = g_F(\rho)F(\rho), \quad (20)$$

$$\psi_G(\rho) = g_G(\rho)G(\rho). \quad (21)$$

The explicit forms of $U_{eff}(\rho)$ and $g(\rho)$ are presented in Appendixes A, B.

Equations (17), (18) are transformed into each other at $\varepsilon \to -\varepsilon, \kappa \to -\kappa$. It follows therefrom that equations (17), (18) describe the motion of particles and antiparticles. In this paper, equation (17) is used for particles for the function $\psi_F(\rho)$ with the effective potential $U_{eff}^F$. The basis for that can be the nonrelativistic limit of the Dirac equation with the lower spinor proportional to $G(\rho)$ vanishing at the zero momentum of the particle $(\mathbf{p}=0)$. In much the same way, the lower spinor with the function $G(\rho)$ vanishes for a particle under the Foldy-Wouthuysen transformation with any value of the momentum $\mathbf{p}$ [30]. On the contrary, for an antiparticle within the non-relativistic limit $\mathbf{p}=0$ and under the Foldy - Wouthuysen transformation, the upper spinor of the Dirac bispinor wave function proportional to $F(\rho)$ vanishes.

### 3.1 Stationary solution $\varepsilon = 0$

#### 3.1.1 Asymptotics of the effective potential

The effective potential $U_{eff}^{F}(\rho)$ at $\varepsilon = 0$ can be presented as (see Appendix B)

$$U_{eff}^{F}(\rho,\varepsilon=0) = -\frac{3}{8}\frac{\alpha^2}{\rho^2(\rho-2\alpha)^2} + \frac{1}{\rho-2\alpha}\left(-\frac{\alpha}{2\rho^2} + \frac{\rho}{2} + \frac{\kappa^2}{2\rho}\right) + \frac{1}{2}\frac{\kappa}{\rho^{3/2}(\rho-2\alpha)^{1/2}} - \frac{1}{2}. \quad (22)$$

The effective potential is real off the event horizon $(\rho > 2\alpha)$ and complex (due to the last but one summand) under the event horizon $(\rho < 2\alpha)$.

The effective potential is singular on the event horizon $(\rho = 2\alpha)$. The leading singularity of potential (22) on both sides of the event horizon is

$$\left. U_{eff}^{F}(\varepsilon=0)\right|_{\rho\to 2\alpha} = -\frac{3}{32}\frac{1}{(\rho-2\alpha)^2} + \mathrm{O}\left(\frac{1}{|\rho-2\alpha|}\right). \quad (23)$$

Coefficient $3/32 < 1/8$ testifies to absence of the mode of particle "drop" to the event horizon [31], [32] and to possibility of existence of the stationary bound state with $\varepsilon = 0$.

The asymptotics of potential (22) at $\rho \to \infty$ has a classical Newtonian form

$$\left. U_{eff}^{F}(\varepsilon=0)\right|_{\rho\to\infty} = \frac{\alpha}{\rho}. \quad (24)$$

At $\rho \to 0$, the asymptotics of the effective potential has the form of a repulsive barrier:

$$\left. U_{eff}^{F}(\varepsilon=0)\right|_{\rho\to 0} = \frac{5}{32}\frac{1}{\rho^2}. \quad (25)$$

#### 3.1.2 Square integrability of radial wave functions

Let us consider the problem of square integrability of wave functions using equation (17) as an example. First, let us examine the behavior of the wave function of equation (17) in the neighborhood of the event horizon of $\rho \to 2\alpha$. Let

$$\left. \psi_F(\rho)\right|_{\rho\to 2\alpha} = \nu^s \sum_{k=0}^{\infty} \varphi_k \nu^k, \quad (26)$$

where $\nu = |\rho - 2\alpha|$.

The indicial equation for (17), taking into account (23), is

$$s(s-1)+\frac{3}{16}=0. \tag{27}$$

Solutions (27) are $s_1 = 3/4, s_2 = 1/4$.

Both the solutions lead to regular square integrable solutions for the wave function $\psi_F(\rho, \varepsilon = 0)$. For an unambiguous selection of the solution, let us turn to asymptotics (14) for the radial function of the Dirac equation $F(\rho)$ with $s = -1/2$ at $\varepsilon = 0$ and to transformation (20)

$$g_F\big|_{\rho \to 2\alpha} \sim \rho^{3/4}. \tag{28}$$

As a result, we obtain

$$\psi_F(\varepsilon = 0)\big|_{\rho \to 2\alpha} = C_3 |\rho - 2\alpha|^{1/4}. \tag{29}$$

Asymptotics (29) corresponds to representation (29) with the solution of indicial equation (27) $s_2 = 1/4$.

Likewise, the consideration at $\rho \to \infty$, taking into account asymptotics (13) and $g_F\big|_{\rho \to \infty} \sim \rho$, leads to the asymptotics

$$\psi_F\big|_{\rho \to \infty} = C_1 \varphi_1(\rho) \rho e^{-\rho\sqrt{1-\varepsilon^2}}. \tag{30}$$

Formally, at $\rho \to 0$, the indicial equation for (17), taking into account asymptotics (25), leads to two solution of $s_1 = 5/4, s_2 = -1/4$. Both the solutions correspond to the square integrable wave functions $\psi_F(\varepsilon = 0)\big|_{\rho \to 0}$.

### 3.1.3 The domain of the radial wave function $\psi_F(\rho)$

Our consideration lead to splitting of the initial domain $\psi_F(\rho)$: $\rho \in (0, \infty)$ into two domains:

$$\rho \in [2\alpha, \infty), \tag{31}$$

$$\rho \in (0, 2\alpha]. \tag{32}$$

Domains (31), (32) on the event horizon $\rho = 2\alpha$ are separated from each other by infinitely deep potential wells $\sim -\frac{3}{32}\frac{1}{(\rho - 2\alpha)^2}$ (see (23)). The wave functions on the event horizon are equal to zero (see (29)). In domain (31), effective potential (22) and the radial wave function $\psi_F(\rho)$ are real. In domain (32), on the contrary, these values become complex. In the domain under the event horizon, there is no possibility to formulate a boundary problem of existence of stationary bound states of particles with a half-spin due to presence of two solutions to the indicial equation for (17) at $\rho \to 0$. Both the solutions correspond to square-integrable solutions for the wave function $\psi_F(\varepsilon=0)\big|_{\rho \to 0}$.

Below, we will examine domain (31) with $\rho \geq 2\alpha$. For solution of the second-order equation $\varepsilon = 0$ there exist regular real square integrable radial wave functions $\psi_F(\varepsilon=0,\rho)$ vanishing at $\rho = 2\alpha$. Below, we will demonstrate this by numerical solutions to equation (17).

## 4. Numerical solutions of the second-order equation

Let us use the Prüfer transformation [25] - [28] to solve equation (17)

### 4.1 Prüfer transformation

Let

$$\psi_F(\rho) = P(\rho)\sin\Phi(\rho),$$
$$\frac{d\psi_F(\rho)}{d\rho} = P(\rho)\cos\Phi(\rho). \tag{33}$$

Then,

$$\psi_F(\rho)\bigg/\frac{\psi_F(\rho)}{d\rho} = \mathrm{tg}\Phi(\rho) \tag{34}$$

and equation (17) can be written as a system of the first-order nonlinear differential equations

$$\frac{d\Phi}{d\rho} = \cos^2\Phi + 2\left(E_{Schr} - U_{eff}^F\right)\sin^2\Phi, \tag{35}$$

$$\frac{d\ln P}{d\rho} = \left(1 - 2\left(E_{Schr} - U_{eff}^F\right)\right)\sin\Phi\cos\Phi. \tag{36}$$

Let us note that equation (36) should be solved after deriving eigenvalues $\varepsilon_n$ and eigenfunctions $\Phi_n(\rho)$ from equation (35).

### 4.2 Asymptotics of functions $\Phi(\rho), P(\rho)$

#### 4.2.1

For bound states and as $\rho \to \infty$, taking into account (30) and (34), we obtain

$$\mathrm{tg}\Phi\big|_{\rho \to \infty} = -\frac{1}{\sqrt{1-\varepsilon^2}},$$
$$\Phi\big|_{\rho \to \infty} = -\mathrm{arctg}\frac{1}{\sqrt{1-\varepsilon^2}} + k\pi. \tag{37}$$

For exponentially increasing solutions in asymptotics (13), $C_2 \neq 0$ and, taking into account (34),

$$\text{tg}\Phi\big|_{\rho\to\infty} = \frac{1}{\sqrt{1-\varepsilon^2}},$$
$$\Phi\big|_{\rho\to\infty} = \text{arctg}\frac{1}{\sqrt{1-\varepsilon^2}} + k\pi, \tag{38}$$

where $k = 0, \pm 1, \pm 2, ...$

**4.2.2**

Let at $\rho \to 2\alpha$

$$\Phi\big|_{\rho\to 2\alpha} = k\pi + A(|\rho - 2\alpha|). \tag{39}$$

Then, $\sin\Phi\big|_{\rho\to 2\alpha} \simeq A(|\rho - 2\alpha|), \cos\Phi\big|_{\rho\to 2\alpha} \simeq 1$.

From consistency (39) with equation (35), taking into account the leading singularity

$U_{eff}^F(\varepsilon = 0)\big|_{\rho\to 2\alpha} = -\frac{3}{32}\frac{1}{(\rho - 2\alpha)^2}$ (see (23)), we obtain

$$1 + \frac{3}{16}A^2 = A \tag{40}$$

with the solutions of $A_1 = 4, A_2 = 4/3$.

Then, we integrate equation (36) at $\rho \to 2\alpha$, taking into account the leading singularity of effective potential (23). As a result, taking into account (33),

$$P\big|_{\rho\to 2\alpha} = C_4 \begin{cases} |\rho - 2\alpha|^{-3/4}, A_1 = 4, \\ |\rho - 2\alpha|^{-1/4}, A_2 = 4/3, \end{cases} \tag{41}$$

$$\psi(\varepsilon = 0)\big|_{\rho\to 2\alpha} = C_4 \begin{cases} 4|\rho - 2\alpha|^{1/4}, A_1 = 4, \\ \frac{4}{3}|\rho - 2\alpha|^{3/4}, A_2 = 4/3. \end{cases} \tag{42}$$

The comparison with (29) shows that solutions (39), (41), (42) with solution of equation (40) $A_1 = 4$ and $C_3 = 4C_4$ are appropriate for our consideration.

**4.3 Numerical method to solve equations (35), (36). General properties of the phase functions $\Phi(\rho)$.**

When solving equations (35), (36) numerically, it is necessary to keep in mind that points $\rho = \infty, \rho = 2\alpha$ are singular points. The behavior of the integral curves of equation (35) in the vicinity of the singular points leads to the fact that integration of equations from "right to left" (from $\rho = \rho_{max}$ to $\rho = \rho_{min}$) is the only correct one. In this case, when specifying initial conditions (37), (38) while integrating, the phase function $\Phi(\rho)$ achieves the neighborhood of the event horizon with asymptotics (39) with $A_1 = 4$. Selection of $\rho_{max} = 10^7$ and $\rho_{min} = 2\alpha + 10^{-8}$ ensures good convergence of numerical results.

In this paper, we applied the following numerical method of solving equation (35). For the permissible set of values of $-1 < \varepsilon < 1$, the Cauchy problem is solved numerically with initial conditions (37), (38) at $\rho \to \infty$. To solve the Cauchy problem, we use the fifth-order Runge-Kutta implicit method with step control (the Ehle scheme of the Radau IIA three-stage method [33]). Defining $\varepsilon_n$ spectrum and eigenfunctions $\Phi_n(\rho)$ and integrating equation (36), one can determine functions $P_n(\rho)$ and, taking into account (33), wave functions $(\psi_F)_n(\rho)$. Then, we can determine the probability density of discovering particles at the distance $\rho$ in the spherical layer $d\rho$

$$w(\rho) = P_n^2(\rho) \sin^2 \Phi_n(\rho) \qquad (43)$$

and probability of discovering particles within the range of $[2\alpha, \rho]$

$$W(\rho) = \int_{2\alpha}^{\rho} P_n^2(\rho) \sin^2 \Phi_n(\rho) d\rho. \qquad (44)$$

In problems of defining the $\varepsilon_n$ spectrum, it is reasonable to use the function of $\Phi(\varepsilon, \rho_{min}) = \Phi(\varepsilon)|_{\rho = \rho_{min}}$.

The numerical experiments have shown the presence of the following essential properties of the function $\Phi(\varepsilon, \rho_{min})$ (similar properties of the function $\Phi$ for simpler potentials independent of $\varepsilon$ have been strictly proved in [25] - [27]).

1. The function $\Phi(\varepsilon, \rho_{min})$ is monotonous over $\varepsilon$.

2. In case of existence of bound states with $|\varepsilon| < 1$, the behavior of the function $\Phi(\varepsilon, \rho_{min})$ is stepwise. In problems with well-defined boundary conditions, when the eigenvalue of $\varepsilon_n$ is achieved, the function $\Phi(\varepsilon, \rho_{min})$ varies stepwise by $\pi$.

$$\left[ \Phi(\varepsilon_0 - \Delta\varepsilon, \rho_{min}) - \Phi(\varepsilon_n + \Delta\varepsilon, \rho_{min}) \right]_{\Delta\varepsilon \to 0} = \pm n\pi. \qquad (45)$$

3. In case of absence of bound states, the variation in the function $\Phi(\varepsilon, \rho_{min})$ within the entire range of $|\varepsilon| < 1$ is lower than $\pi$.

# 5. Results of numerical calculations to determine eigenfunctions for solving $\varepsilon = 0$

## 5.1 Analysis of physical acceptability boundaries of the solution $\varepsilon = 0$

The solution of $\varepsilon = 0$ is formally valid for any value of the gravitational coupling constant $\alpha$. However, in order to achieve such a strong coupling of $\varepsilon_b = 1$, rather high values of $\alpha \geq \alpha_{min}$ are necessary. For instance, the energy of a bound electron $\varepsilon_e = 0$ in the Coulomb field of a nucleus with the number of protons $Z$ in the state $1S_{1/2}$ is achieved at the value of the electromagnetic coupling constant of $\alpha_{fs} Z \approx \frac{1}{137} Z \approx 1,06 (Z \approx 140)$. For $2S_{1/2}$ and $3S_{1/2}$ states, there are similar values of $\alpha_{fs} Z \approx 1.42$ and $1.9$ [34].

In order to determine the value of $\alpha_{min}$, let us turn to the stationary regular solution for the Kerr metric (see item 6)

$$\varepsilon_K = \frac{\alpha_a m_\varphi}{2\alpha (\rho_+)_k}, \quad (46)$$

where $(\rho_+)_k = \alpha \pm \sqrt{\alpha^2 - \alpha_a^2}$; $\alpha_a = \frac{a}{l_c} = \frac{J}{Mcl_c}$; $\alpha^2 > \alpha_a^2$.

In absence of rotation, $\alpha_a = 0$, the Kerr metric turns into the Schwarzschild metric with solution (46) of $\varepsilon_S = 0$.

The maximal value of $\alpha_a$ corresponds to the extreme Kerr field $\alpha_a^2 = \alpha^2$. In this case,

$$\varepsilon_K^{extr} = \frac{m_\varphi}{2\alpha}. \quad (47)$$

In order to achieve the maximal energy of bound state near $\varepsilon_{max} \sim 1$ (at the same time, $\varepsilon_{max} < 1$) the minimal value of $\alpha_{min}$ in (47) is

$$\alpha_{min} \simeq \frac{m_\varphi}{2}, \quad m_\varphi > 0. \quad (48)$$

With decrease in $\alpha_a$ from $\alpha_a = \alpha_{min}$ to $\alpha_a = 0$, the energy of bound state in (46) with the value of $\alpha_{min}$ from (48) decreases from $\varepsilon_K^{extr}$ (the extreme Kerr field) to $\varepsilon_S = 0$ (the Schwarzschild field). The minimal value of $\alpha_{min}$ in (48) corresponds to the minimal value of $(m_\varphi)_{min} = 1/2$, i.e., $\alpha_{min} \simeq 0.25$

## 5.2 Results of numerical calculations

Table 1 presents distances $\rho_m$ from maxima of the probability densities up to the event horizons, derived from the calculations for different values of $\alpha, j, l$. Figures 1,2 present normalized densities of probability (43) and integral probabilities (44) for some values of $\alpha, j, l$.

Table 1. Values $\rho_m$ versus $\kappa$ ($j,l$) and $\alpha$.

| $\kappa$ ($j,l$) | $-1$ ($j=1/2, l=0$) | $+1$ ($j=1/2, l=1$) | $-2$ ($j=3/2, l=1$) | $+2$ ($j=3/2, l=2$) | $-3$ ($j=5/2, l=2$) | $+3$ ($j=5/2, l=3$) |
|---|---|---|---|---|---|---|
| $\alpha$ | 0.25 | 0.5 | 0.9 | 1 | 1.4 | 1.5 |
| $\rho_m$ | 0.043 | 0.028 | 0.017 | 0.015 | 0.011 | 0.01 |

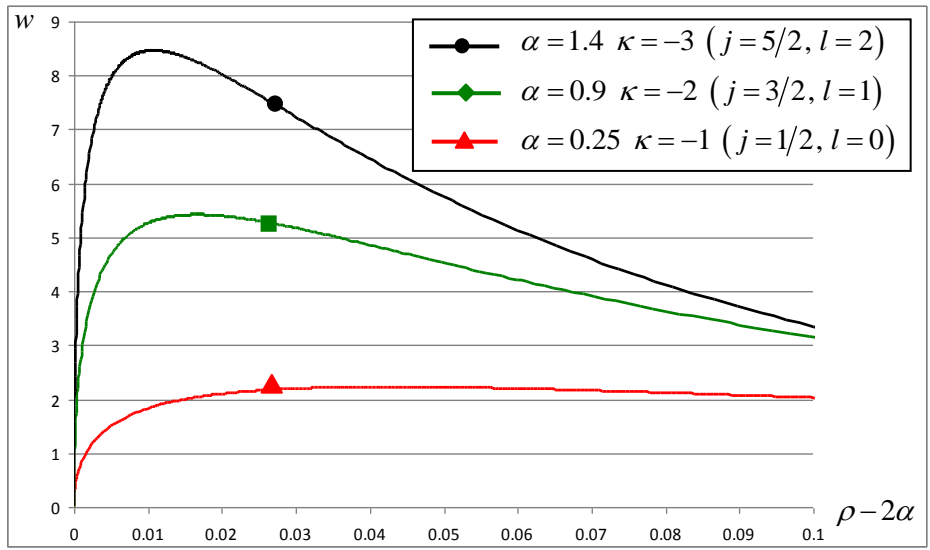

Figure 1. Normalized probability densities $w(\rho-2\alpha)$, $\varepsilon=0$.

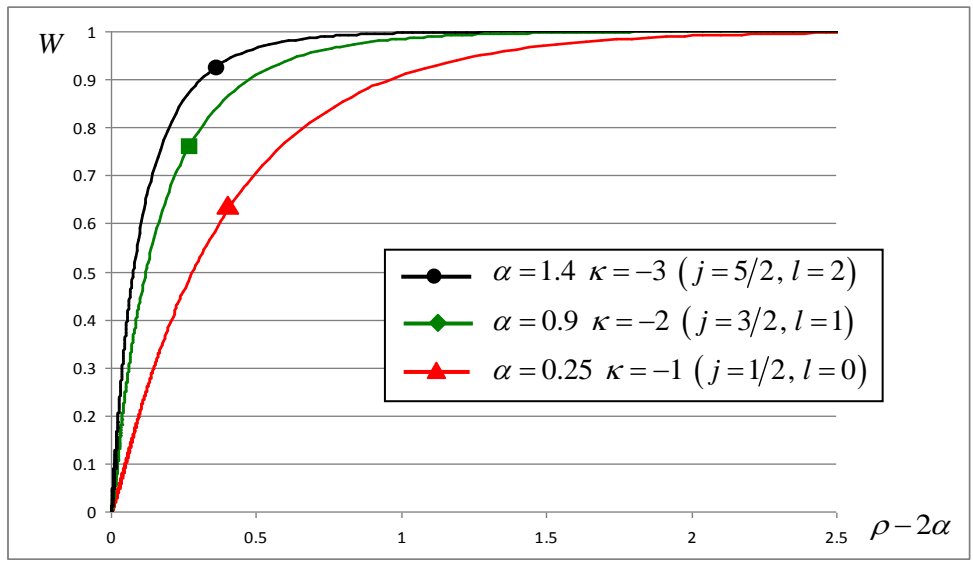

Figure 2. Integral probability $W(\rho-2\alpha)$, $\varepsilon=0$.

It can be seen that unlike the electronic structure of atoms of the Periodic system, localization of particles near the event horizon increases with increase in $j,l$.

For $\alpha_{min} = 0.25$, let us make examples of admissible mass values of a collapsar and a half-spin particle. It is seen from ratio (2) that for an electron with mass $m = m_e \simeq 0.5 MeV$, the minimal possible mass of the collapsar should be equal to $M \approx 10^{17}$ g. For a neutrino with mass $m = m_\nu \approx 1 eV$, the minimal possible mass of the collapsar is $M \approx 0.6 \cdot 10^{25}$ g. Naturally, higher masses of collapsars also ensure achievement of binding $\varepsilon_b = 1$.

## 6. Stationary bound states of half-spin particles in external Reissner-Nordström, Kerr, Kerr-Newman fields

Energies of bound states of Dirac particles are presented below in Schwarzschild, Reissner-Nordström, Kerr, Kerr-Newman fields. The solutions for the latter three fields were obtained methodologically, in the same way as for the Schwarzschild field in this paper.

### 6.1 Schwarzschild field

$$\varepsilon_S = 0. \tag{49}$$

### 6.2 Reissner-Nordström field

$$(\rho_+)_{RN} = \alpha + \sqrt{\alpha^2 - \alpha_Q^2}; \quad (\rho_-)_{RN} = \alpha - \sqrt{\alpha^2 - \alpha_Q^2}; \quad \alpha^2 > \alpha_Q^2$$

$$\varepsilon_{R-N} = \alpha_{em}/(\rho_+)_{RN}, \quad \rho \in [\rho_+, \infty], \tag{50}$$

$$\varepsilon_{R-N} = \alpha_{em}/(\rho_-)_{RN}, \quad \rho \in (0, \rho_-]. \tag{51}$$

### 6.3 Kerr field

$$(\rho_+)_K = \alpha + \sqrt{\alpha^2 - \alpha_a^2}; \quad (\rho_-)_K = \alpha - \sqrt{\alpha^2 - \alpha_a^2}; \quad \alpha^2 > \alpha_a^2,$$

$$\varepsilon_K = \frac{m_\varphi \alpha_a}{\alpha_a^2 + (\rho_+)_K^2}, \quad \rho \in [\rho_+, \infty), \tag{52}$$

$$\varepsilon_K = \frac{m_\varphi \alpha_a}{\alpha_a^2 + (\rho_-)_K^2}, \quad \rho \in (0, \rho_-]. \tag{53}$$

### 6.4 Kerr-Newman field

$$(\rho_+)_{KN} = \alpha + \sqrt{\alpha^2 - \alpha_a^2 - \alpha_Q^2}; \quad (\rho_-)_{KN} = \alpha - \sqrt{\alpha^2 - \alpha_a^2 - \alpha_Q^2}; \quad \alpha^2 > \alpha_a^2 + \alpha_Q^2,$$

$$\varepsilon_{K-N} = \frac{m_\varphi \alpha_a + \alpha_{em}(\rho_+)_{KN}}{\alpha_a^2 + (\rho_+)_{KN}^2}, \quad \rho \in [\rho_+, \infty), \tag{54}$$

$$\varepsilon_{K-N} = \frac{m_\varphi \alpha_a + \alpha_{em}(\rho_-)_{KN}}{\alpha_a^2 + (\rho_-)_{KN}^2}, \quad \rho \in (0, \rho_-]. \tag{55}$$

In (50) - (55), the new denotations are introduced:

$$\alpha_Q = r_Q/l_c, \quad r_Q = \sqrt{G}Q/c^2, \quad \alpha_{em} = eQ/\hbar c, \quad \alpha_a = a/l_c, \quad a = J/Mc.$$

Everywhere, the permissible energy interval of a particle in the bound state is the following one:

$$-1 < \varepsilon < 1. \tag{56}$$

For each of solutions (50) - (55), there exists a square-integrable wave eigenfunction being a solution to the Schrödinger-type equation with the effective potential. The said solutions are to be validated in our next papers. The intervals of variations in $\alpha, \alpha_Q, \alpha_{em}, \alpha_a, m_\varphi$, at which the existence of stationary bound states of fermions under investigation is possible, are to be specified ibidem.

## 7. Stationary bound states and particles of dark matter

Atomic systems with stationary bound states of half-spin particles in the field of collapsars determined by solutions (49) - (55) are good candidates for the role of dark matter particles. Indeed, let us consider, for instance, solution (49) $\varepsilon_S = 0$ for the Schwarzschild field. In this case for the Schwarzschild collapsar with mass $M$, an atomic system with bound half-spin particle with $\varepsilon_S = 0$ is possible.

The atomic system, the Schwarzschild collapsar with uncharged Dirac particle with $\varepsilon_S = 0$, interacts with other objects only gravitationally. Due to the absence of quantum transitions between the states with different $j, l$, such system neither emits nor absorbs light or other kinds of radiation. This system can be detected only through gravitational interaction. Masses of such systems as $\alpha \geq \alpha_{min}$ are restricted by relation $Mm/M_P^2 \geq \alpha_{min}$ (see (2)) and should be chosen from the condition of the best agreement with the Standard cosmological model.

Other solutions (50) - (55) to the problem of dark matter will be discussed in future papers.

## 8. Transition from quantum mechanics to classical description

A quantum mechanics of nonevaporating collapsars and black holes near the event horizons has no classical analog because of objects under consideration are demonstration of extremely strong gravitational fields. Both the bound states of fermions with $\varepsilon = 0$ and localization near the event horizon of collapsars, and the resonant state of fermions decaying with time for black holes [8] - [13] on its own cannot have the classical analogs. However, the considered in section 7 atomic systems with sufficiently great numbers of bound half-spin particles can lead to classical description. Let us consider an atomic system with mass of the Schwarzschild collapsar $M$, bound fermion with mass $m_1$, $\varepsilon_1 = 0$ and gravitational coupling constant $\alpha_1$. The next fermion with mass $m_2$ will be in the gravitational fields generated by the Schwarzschild metric with mass $M$ and fermion with mass $m_1$. Effective coupling constant for the mass $m_2$ will be bigger then primarily for the mass $m_1$ ($\alpha_2 > \alpha_1$) and the second fermion can be also connected with energy $\varepsilon_2 = 0$. This process can be continued and in case of $\sum m_i \gg M$ in conditions of a self-consistent gravitational field there will actually be no effect of the event horizon on the new fermions. In this case the system under consideration can be described by classical celestial mechanics and the closed Keplerian orbits will exist for test particles.

## 9. Conclusions

The existence of the stationary solution to the second-order equation has been proved in the paper for half-spin particles in the Schwarzschild field, within the applicability range of one-particle quantum mechanics. The solution corresponds to the energy of state $E = 0$ (the binding energy is $E_b = mc^2$). For each of the values of quantum numbers $j, l$, the solution is implemented at the value of gravitational coupling constant $\alpha > 0.25$. The bound particles with $E = 0$ are, with the overwhelming probability, at the distance from the event horizon within the range from zero to several fractions of the Compton wave length of a fermion depending on the values of $\alpha, j, l$.

The singularity of the effective potential in the neighborhood of the event horizon at $\varepsilon_S = 0$ $\sim -\dfrac{3}{32}\dfrac{1}{(\rho - \rho_+)^2}$ allows existence of bound states of fermions. To solution $\varepsilon_S = 0$, it is important whether this singularity of the potential is preserved under coordinate transformation of the Schwarzschild metric to other static stationary metrics proposed in due time with the view to eliminate the coordinate singularity of the initial metric on the event horizon.

In [35], conservation of the singularity of effective potential (23) has been proved for the static Schwarzschild metric in the isotropic coordinates [36], for the stationary Eddington-Finkelstein metric [37], [38], for the stationary Painleve-Gullstrand metric [39], [40].

In the paper, the attention is drawn to the possibility of using in the Standard cosmological model the Schwarzschild collapsars with fermions, being in bound states with $\varepsilon_S = 0$ as particles of the dark matter. The atomic systems of this a type neither absorb nor emit light or other kinds of radiation and interact with the environment only gravitationally.

As the result, our consideration has shown that use of the self-conjugate second-order equation extends possibilities of obtaining regular solutions of quantum mechanics of half-spin particle motion in external gravitational and electromagnetic fields.

**Acknowledgements**

APPENDIX A

## Self-conjugate second-order equation for half-spin particles in the gravitational Schwarzschild field

Dirac equation (6) with Hamiltonian (5) for the stationary states of $\psi_\eta(t,\rho,\theta,\varphi) = e^{-i\varepsilon t}\psi_\eta(\rho,\theta,\varphi)$ can be written as

$$(\varepsilon - H_\eta)\psi_\eta(\rho,\theta,\varphi) = 0. \tag{A.1}$$

Let us multiply equality on the left (A.1) by the operator $(\varepsilon + H_\eta)$

$$(\varepsilon + H_\eta)(\varepsilon - H_\eta)\psi_\eta(\rho,\theta,\varphi) = 0. \tag{A.2}$$

Taking into account (5), we obtain

$$\left\{\varepsilon^2 - f_S + \left(f_S\frac{\partial}{\partial\rho} + \frac{1}{\rho} - \frac{\alpha}{\rho^2}\right)\left(f_S\frac{\partial}{\partial\rho} + \frac{1}{\rho} - \frac{\alpha}{\rho^2}\right) + \frac{f_S}{\rho^2}\left[\left(\frac{\partial}{\partial\theta} + \frac{1}{2}\mathrm{ctg}\theta\right)\times\right.\right.$$
$$\left.\times\left(\frac{\partial}{\partial\theta} + \frac{1}{2}\mathrm{ctg}\theta\right) + \frac{1}{\sin^2\theta}\frac{\partial^2}{\partial\varphi^2} + i\Sigma^{\underline{3}}\frac{\partial}{\partial\theta}\left(\frac{1}{\sin\theta}\right)\frac{\partial}{\partial\varphi}\right] - i\gamma^0\gamma^{\underline{3}}f_S\frac{d}{d\rho}\left(\sqrt{f_S}\right) + $$
$$\left. + f_S\frac{d}{d\rho}\left(\sqrt{f_S}\frac{1}{\rho}\right)\left[i\Sigma^{\underline{3}}\left(\frac{\partial}{\partial\theta} + \frac{1}{2}\mathrm{ctg}\theta\right) - i\Sigma^{\underline{1}}\frac{1}{\sin\theta}\frac{\partial}{\partial\varphi}\right]\right\}\psi_\eta(\rho,\theta,\varphi) = 0. \tag{A.3}$$

In (A.3), as well as earlier in item 2.1, there was an equivalent substitution of matrices (9);

$$\Sigma^{\underline{k}} = \begin{pmatrix} \sigma^k & 0 \\ 0 & \sigma^k \end{pmatrix}.$$

Dirac equations for upper and lower components of the bispinor

$$\psi_\eta(\rho,\theta,\varphi,t) = \begin{pmatrix} U(\rho,\theta,\varphi) \\ W(\rho,\theta,\varphi) \end{pmatrix} e^{-i\varepsilon t} \tag{A.4}$$

have the form

$$\left(\varepsilon-\sqrt{f_s}\right)U=\left(-i\sigma^3\left(f_s\frac{\partial}{\partial\rho}+\frac{1}{\rho}-\frac{\alpha}{\rho^2}\right)-i\sigma^1\sqrt{f_s}\frac{1}{\rho}\left(\frac{\partial}{\partial\theta}+\frac{1}{2}\operatorname{ctg}\theta\right)-i\sigma^2\sqrt{f_s}\frac{1}{\rho\sin\theta}\frac{\partial}{\partial\varphi}\right)W,$$
$$\left(\varepsilon+\sqrt{f_s}\right)W=\left(-i\sigma^3\left(f_s\frac{\partial}{\partial\rho}+\frac{1}{\rho}-\frac{\alpha}{\rho^2}\right)-i\sigma^1\sqrt{f_s}\frac{1}{\rho}\left(\frac{\partial}{\partial\theta}+\frac{1}{2}\operatorname{ctg}\theta\right)-i\sigma^2\sqrt{f_s}\frac{1}{\rho\sin\theta}\frac{\partial}{\partial\varphi}\right)U.$$
(A.5)

As a result, taking into account (A.5), equation (A.3) can be written for one of the spinors $U(\rho,\theta,\varphi)$ or $W(\rho,\theta,\varphi)$. For the spinor $U(\rho,\theta,\varphi)$, equation (A.3) has the view

$$\left\{\varepsilon^2-f_s+\left(f_s\frac{\partial}{\partial\rho}+\frac{1}{\rho}-\frac{\alpha}{\rho^2}\right)^2+\frac{f_s}{\rho^2}\left[\left(\frac{\partial}{\partial\theta}+\frac{1}{2}\operatorname{ctg}\theta\right)^2+\frac{1}{\sin^2\theta}\frac{\partial^2}{\partial\varphi^2}+i\sigma^3\frac{\partial}{\partial\theta}\left(\frac{1}{\sin\theta}\right)\frac{\partial}{\partial\varphi}\right]+\right.$$
$$+f_s\frac{d}{d\rho}\left(\sqrt{f_s}\frac{1}{\rho}\right)\left[i\sigma^2\left(\frac{\partial}{\partial\theta}+\frac{1}{2}\operatorname{ctg}\theta\right)-i\sigma^1\frac{1}{\sin\theta}\frac{\partial}{\partial\varphi}\right]+f_s\frac{d}{d\rho}\left(\sqrt{f_s}\right)\frac{1}{\varepsilon+\sqrt{f_s}}\times$$
$$\times\left[-f_s\frac{\partial}{\partial\rho}-\frac{1}{\rho}+\frac{\alpha}{\rho^2}-i\sigma^2\sqrt{f_s}\frac{1}{\rho}\left(\frac{\partial}{\partial\theta}+\frac{1}{2}\operatorname{ctg}\theta\right)+i\sigma^1\sqrt{f_s}\frac{1}{\rho\sin\theta}\frac{\partial}{\partial\varphi}\right]\right\}U(\rho,\theta,\varphi)=0.$$
(A.6)

Then, the variables can be separated. From representation (7) it follows that

$$U(r,\theta,\varphi)=F(\rho)\xi(\theta)e^{im_\varphi\varphi}.$$
(A.7)

Using Brill-Wheeler equation (8) and its squared representation [41]

$$\left[\left(\frac{\partial}{\partial\theta}+\frac{1}{2}\operatorname{ctg}\theta\right)^2+\frac{1}{\sin^2\theta}\frac{\partial^2}{\partial\varphi^2}+i\sigma^3\frac{\partial}{\partial\theta}\left(\frac{1}{\sin\theta}\right)\frac{\partial}{\partial\varphi}\right]\xi(\theta)e^{im_\varphi\varphi}=-\kappa^2\xi(\theta)e^{im_\varphi\varphi},$$
(A.8)

we can derive the second-order equation for the radial function $F(\rho)$

$$\left\{\varepsilon^2-f_s+\left(f_s\frac{\partial}{\partial\rho}+\frac{1}{\rho}-\frac{\alpha}{\rho^2}\right)^2-\frac{f_s\kappa^2}{\rho^2}+f_s\kappa\frac{d}{d\rho}\left(\sqrt{f_s}\frac{1}{\rho}\right)-\right.$$
$$\left.-f_s\frac{d}{d\rho}\left(\sqrt{f_s}\right)\frac{1}{\varepsilon+\sqrt{f_s}}\frac{\kappa\sqrt{f_s}}{\rho}-f_s\frac{d}{d\rho}\left(\sqrt{f_s}\right)\frac{1}{\varepsilon+\sqrt{f_s}}\left(f_s\frac{\partial}{\partial\rho}+\frac{1}{\rho}-\frac{\alpha}{\rho^2}\right)\right\}F(\rho)=0.$$
(A.9)

In equation (A.9), the third and the last summands are not self-conjugate. For self-conjugacy (A.9), let us perform nonunitary transformation of the similarity

$$F(\rho)=g_F^{-1}(\rho)\psi_F(\rho).$$
(A.10)

If in equation (12), we denote

$$A(\rho)=-\frac{1}{f_s}\left(\frac{1+\kappa\sqrt{f_s}}{\rho}-\frac{\alpha}{\rho^2}\right),$$
(A.11)

$$B(\rho)=\frac{1}{f_s}\left(\varepsilon+\sqrt{f_s}\right),$$
(A.12)

$$C(\rho)=-\frac{1}{f_s}\left(\varepsilon-\sqrt{f_s}\right),$$
(A.13)

$$D(\rho) = -\frac{1}{f_S}\left(\frac{1-\kappa\sqrt{f_S}}{\rho} - \frac{\alpha}{\rho^2}\right) \quad (A.14)$$

and, besides, introduce denotations of

$$A_F(\rho) = -\frac{1}{B}\frac{dB}{d\rho} - A - D, \quad (A.15)$$

$$A_G(\rho) = -\frac{1}{C}\frac{dC}{d\rho} - A - D, \quad (A.16)$$

then, the desired transformation is

$$g_F(\rho) = \exp\frac{1}{2}\int A_F(\rho')d\rho'. \quad (A.17)$$

As a result, if we write equation (A.9) as

$$\hat{M}F(\rho) = 0,$$

then the transformed self-conjugate equation has the form

$$g_F(\rho)\hat{M}g_F^{-1}\psi_F(\rho) = 0 \quad (A.18)$$

Equation (A.18) can be written as the Schrödinger-type second-order equation with the effective potential $U_{eff}^F(\rho)$

$$\frac{d^2\psi_F}{d\rho^2} + 2\left(E_{Schr} - U_{eff}^F\right)\psi_F = 0, \quad (A.19)$$

where

$$E_{Schr} = \frac{1}{2}\left(\varepsilon^2 - 1\right), \quad (A.20)$$

$$U_{eff}^F = -\frac{1}{4}\frac{1}{B}\frac{d^2B}{d\rho^2} + \frac{3}{8}\left(\frac{1}{B}\frac{dB}{d\rho}\right)^2 - \frac{1}{4}(A-D)\frac{1}{B}\frac{dB}{d\rho} + \frac{1}{4}\frac{d}{d\rho}(A-D) + \\ + \frac{1}{8}(A-D)^2 + \frac{1}{2}BC + E_{Schr}. \quad (A.21)$$

In equation (A.19), summand $E_{Schr}$ (A.20) is selected and simultaneously added to (A.21). On the one hand, it is done for equation (A.19) to have the form of the Schrödinger-type equation and, on the other hand, to ensure classical asymptotics of the effective potential at $\rho \to \infty$.

For the lower spinor $W(\rho,\theta,\varphi)$ with the radial function $G(\rho)$, the appropriate equations have the form

$$G(\rho) = g_G^{-1}\psi_G(\rho), \quad (A.22)$$

$$g_G(\rho) = \exp\frac{1}{2}\int A_G(\rho')d\rho', \qquad (A.23)$$

$$\frac{d^2\psi_G}{d\rho^2} + 2\left(E_{Schr} - U_{eff}^G\right)\psi_G = 0, \qquad (A.24)$$

$$U_{eff}^G = -\frac{1}{4}\frac{1}{C}\frac{d^2C}{d\rho^2} + \frac{3}{8}\left(\frac{1}{C}\frac{dC}{d\rho}\right)^2 + \frac{1}{4}\frac{(A-D)}{C}\frac{dC}{d\rho} - \frac{1}{4}\frac{d}{d\rho}(A-D) +$$
$$+\frac{1}{8}(A-D)^2 + \frac{1}{2}BC + E_{Schr}. \qquad (A.25)$$

# APPENDIX B

## Effective potential of the Schwarzschild field in the Schrödinger-type equation

In compliance with (A.11) - (A.14), (A.21), we can obtain

$$\frac{3}{8}\frac{1}{B^2}\left(\frac{dB}{d\rho}\right)^2 = \frac{3}{8}\left(\frac{2\alpha}{\rho(\rho-2\alpha)} - \frac{\alpha}{\varepsilon\rho(\rho(\rho-2\alpha))^{1/2} + \rho(\rho-2\alpha)}\right)^2, \qquad (B.1)$$

$$-\frac{1}{4}\frac{1}{B}\frac{d^2B}{d\rho^2} = -\frac{2\alpha^2}{\rho^2(\rho-2\alpha)^2} - \frac{\alpha}{\rho^2(\rho-2\alpha)} + \frac{5}{4}\frac{\alpha^2}{\varepsilon\rho^{5/2}(\rho-2\alpha)^{3/2} + \rho^2(\rho-2\alpha)^2} +$$
$$+\frac{\alpha}{2\left[\varepsilon\rho^{5/2}(\rho-2\alpha)^{1/2} + \rho^2(\rho-2\alpha)\right]}, \qquad (B.2)$$

$$\frac{1}{4}\frac{d}{d\rho}(A-D) = \frac{\kappa(\rho-\alpha)}{2\rho^{3/2}(\rho-2\alpha)^{3/2}}, \qquad (B.3)$$

$$-\frac{1}{4}\frac{(A-D)}{B}\frac{dB}{d\rho} = -\frac{\alpha\kappa}{\rho^{3/2}(\rho-2\alpha)^{3/2}} + \frac{\alpha\kappa}{2\left[\varepsilon\rho^2(\rho-2\alpha) + \rho^{3/2}(\rho-2\alpha)^{3/2}\right]}, \qquad (B.4)$$

$$\frac{1}{8}(A-D)^2 = \frac{\kappa^2}{2\rho(\rho-2\alpha)}, \qquad (B.5)$$

$$\frac{1}{2}BC = -\frac{1}{2}\frac{\rho^2\varepsilon^2}{(\rho-2\alpha)^2} + \frac{1}{2}\frac{\rho}{\rho-2\alpha}. \qquad (B.6)$$

The sum of expressions of $E_{Schr}$ and (B.1) - (B.6), leads to the desired expression for the effective potential $U_{eff}^F$.

The asymptotics is

$$U_{eff}^F(\varepsilon=0)\Big|_{\rho\to 2\alpha} = -\frac{3}{32(\rho-2\alpha)^2}. \qquad (B.7)$$